\documentclass[aps,prb,twocolumn,showpacs]{revtex4-1}

\begin{document}

\title{Nonlinear optical properties of undoped and doped with Zr and Nb
KTiOPO$_4$ crystals}

\author{A. I. Lebedev}
\email[]{swan@scon155.phys.msu.ru}
\affiliation{Faculty of Physics, Moscow State University, Moscow, 119991 Russia}

\date{\today}

\begin{abstract}
The structure of the ferroelectric phase of undoped KTiOPO$_4$ and its solid
solutions with zirconium and niobium is studied from first principles within
the density functional theory. The second-order nonlinear susceptibility tensor
and the spontaneous polarization of these materials are obtained. It is shown
that an improvement in nonlinear optical properties of KTiOPO$_4$ upon its
doping with Zr and Nb cannot be explained by a systematic change in the
composition of crystals and is apparently associated with the occurrence of
defects. Possible structures of such defects are discussed.

\texttt{DOI: 10.3103/S1062873816090276}
\end{abstract}

\pacs{42.65.An, 42.70.Mp, 77.84.-s}

\maketitle

High nonlinear optical properties of ferroelectrics are widely used
in different optical devices, such as parametric amplifiers, frequency
mixers and multipliers. Among the materials for nonlinear optics,
ferroelectric potassium titanyl phosphate KTiOPO$_4$ (KTP) holds
a special place. This material is characterized by high optical nonlinearity
and optical damage resistance, weak absorption in the visible and IR
spectral regions, and good mechanical properties. The search for ways to
improve KTP performance shows that doping it with zirconium and niobium
results in considerable enhancement of its nonlinear optical
properties.~\cite{SolidStateCommun.73.97,JCrystGrowth.171.472,InorgMater.40.1321,
CrystallogrRep.53.678}
However, the mechanism of the doping effect on these properties remains
unclear. In this work, to understand if the improvement in the nonlinear
optical properties of KTP is a consequence of systematic changes in the
crystal composition, we performed the first-principles calculations of the
structure of the ferroelectric phase for undoped KTP and its solid solutions
with zirconium and niobium and studied the changes that occur in the
second-order nonlinear susceptibility tensor upon doping these crystals.

Calculations were performed within the density functional theory using the
\texttt{ABINIT} program on 64-atom unit cells of KTP. Pseudopotentials for
the K, Ti, Zr, Nb, and O atoms used in our calculations were taken from
Refs.~\onlinecite{PhysSolidState.51.362,PhysSolidState.52.1448}.  The
nonrelativistic pseudopotential for the P atom (electronic configuration
$3s^{0.9}3p^{0.1}4s^0$) was constructed according to the
scheme,~\cite{PhysRevB.41.1227} using the \texttt{OPIUM} program with the
parameters $r_s = r_p = 1.36$, $q_s = q_p = 7.07$, $r_{\rm min} = 0.01$,
$r_{\rm max} = 1.28$, and $V_{\rm loc} = 5.0$~a.u. (for explanation of
the parameters see Ref.~\onlinecite{PhysSolidState.51.362}).
We used the local density approximation (LDA) for the exchange-correlation
energy functional. The maximum energy of plane waves was 816~eV
and the integration over the Brillouin zone was done on a 4$\times$6$\times$4
Monkhorst--Pack mesh. Optimization of the lattice parameters and
the positions of atoms was continued until the residual forces acting on the
atoms were less than 0.5~meV/{\AA}. Polarization was calculated using the Berry
phase method. The second-order nonlinear susceptibility tensor was calculated
within the density functional perturbation theory.~\cite{PhysRevB.71.125107}

\begin{table*}
\caption{\label{table1}Comparison of calculated and experimental lattice
parameters and positions of atoms in the ferroelectric $Pna2_1$ phase of
potassium titanyl phosphate.}
\begin{ruledtabular}
\begin{tabular}{cccccccc}
Atom & Position & \multicolumn{3}{c}{This work} & \multicolumn{3}{c}{Experiment (Ref.~\onlinecite{ZKrist.139.103})} \\
&& $x$ & $y$ & $z$ & $x$ & $y$ & $z$ \\
\hline
$a$  && \multicolumn{3}{c}{12.8313~{\AA}} & \multicolumn{3}{c}{12.814~{\AA}} \\
$b$  && \multicolumn{3}{c}{ 6.4155~{\AA}} & \multicolumn{3}{c}{ 6.404~{\AA}} \\
$c$  && \multicolumn{3}{c}{10.6294~{\AA}} & \multicolumn{3}{c}{10.616~{\AA}} \\
\hline
Ti(1) & $4a$ & 0.37331 & 0.50212 & 0.00000 & 0.3729 & 0.5001 & 0.0000 \\
Ti(2) & $4a$ & 0.24443 & 0.26442 & 0.24883 & 0.2467 & 0.2692 & 0.2515 \\
P(1)  & $4a$ & 0.49669 & 0.33802 & 0.25625 & 0.4980 & 0.3364 & 0.2608 \\
P(2)  & $4a$ & 0.18022 & 0.50467 & 0.50730 & 0.1808 & 0.5020 & 0.5131 \\
K(1)  & $4a$ & 0.36995 & 0.77783 & 0.29618 & 0.3780 & 0.7804 & 0.3120 \\
K(2)  & $4a$ & 0.10842 & 0.70554 & 0.05173 & 0.1053 & 0.6989 & 0.0669 \\
O(1)  & $4a$ & 0.27734 & 0.55220 & $-$0.11732 & 0.2762 & 0.5405 & $-$0.1093 \\
O(2)  & $4a$ & 0.27739 & 0.46732 & 0.13565 & 0.2753 & 0.4667 & 0.1447 \\
O(3)  & $4a$ & 0.48691 & 0.49776 & 0.14462 & 0.4862 & 0.4865 & 0.1504 \\
O(4)  & $4a$ & 0.50533 & 0.46807 & 0.38368 & 0.5100 & 0.4650 & 0.3836 \\
O(5)  & $4a$ & 0.39749 & 0.19140 & 0.27350 & 0.3999 & 0.1989 & 0.2800 \\
O(6)  & $4a$ & 0.59452 & 0.18859 & 0.23490 & 0.5937 & 0.1925 & 0.2422 \\
O(7)  & $4a$ & 0.10757 & 0.31141 & 0.53695 & 0.1120 & 0.3105 & 0.5421 \\
O(8)  & $4a$ & 0.10806 & 0.69928 & 0.48227 & 0.1111 & 0.6918 & 0.4878 \\
O(9)  & $4a$ & 0.25629 & 0.53422 & 0.62514 & 0.2526 & 0.5396 & 0.6285 \\
O(10) & $4a$ & 0.25505 & 0.46795 & 0.39037 & 0.2530 & 0.4605 & 0.3995 \\
\end{tabular}
\end{ruledtabular}
\end{table*}

The calculated lattice parameters and the positions of atoms in the
ferroelectric $Pna2_1$ phase of undoped potassium titanyl phosphate
are compared with the experimental data~\cite{ZKrist.139.103} in
Table~\ref{table1}. It is seen that the calculations reproduce the
structure of the polar phase quite well. Calculations of spontaneous
polarization for the obtained structure give a value of $P_s = 0.259$~C/m$^2$,
which is in good agreement with the experimental data (0.237~C/m$^2$
according to Ref.~\onlinecite{ApplPhysLett.73.3650} and 0.200~C/m$^2$
according to Ref.~\onlinecite{PhysRevB.66.094102}). The calculated values
of five non-zero independent components of the second-order nonlinear
susceptibility tensor $d_{i \nu}$ (whose values $d_{31}$, $d_{15}$, $d_{32}$,
and $d_{24}$ coincide pairwise in the low-frequency limit) for undoped KTP
are compared with the experimental
data~\cite{JApplPhys.47.4980,OptLett.17.982,JOptSocAmB.11.750,ApplOpt.36.5902,
JOptSocAmB.14.2268,ApplOpt.43.3319}
in Table~\ref{table2}. It is seen that if one takes into account the spread
of the experimental data, the agreement is quite good.

In calculating the properties of doped
samples, one of eight titanium atoms in the unit cell was replaced
with an impurity atom. The symmetry of the unit cell was thus reduced
to $P1$. A comparison of the energies of the structures in which zirconium
atom substitutes for titanium atoms at positions Ti(1) and Ti(2) shows
that Zr predominantly occupies position Ti(2) (the energy of the 64-atom
unit cell with Zr atom at this position is 155~meV lower than that of
the structure with Zr atom at position Ti(1)).%
    \footnote{In this work, we do not show tables with the coordinates of all
    64 atoms and the lattice parameters for the obtained unit cells because
    of their large size.}
Our conclusion about the preferential location of Zr at position Ti(2) in KTP
is consistent with the experimental data.~\cite{CrystallogrRep.52.659,CrystallogrRep.54.219}

\begin{table*}
\caption{\label{table2}Components of the second-order nonlinear susceptibility
tensor $d_{i \nu}$ in undoped KTP crystals (in pm/V).}
\begin{ruledtabular}
\begin{tabular}{cccccccc}
Component & This work & \multicolumn{6}{c}{Experimental data} \\
&& Ref.~\onlinecite{JApplPhys.47.4980} & Ref.~\onlinecite{OptLett.17.982} & Ref.~\onlinecite{JOptSocAmB.11.750} & Ref.~\onlinecite{ApplOpt.36.5902} & Ref.~\onlinecite{JOptSocAmB.14.2268} & Ref.~\onlinecite{ApplOpt.43.3319} \\
\hline
$d_{31}$  &  1.19     & 6.5  & 2.54 &  1.40 & ---  &  2.2 &  2.12 \\
$d_{32}$  &  4.51     & 5.0  & 4.35 &  2.65 & ---  &  3.7 &  3.75 \\
$d_{33}$  & 15.06     & 13.7 & 16.9 & 10.70 & 17.4 & 14.6 & 15.4 \\
$d_{24}$  &  4.51     &  7.6 & 3.64 &  2.65 & 3.37 &  3.7 & --- \\
$d_{15}$  &  1.19     &  6.1 & 1.91 &  1.40 & 1.78 &  1.9 & 2.02 \\
\end{tabular}
\end{ruledtabular}
\end{table*}

In calculating the properties of niobium-doped KTP, the potassium atom closest
to the Nb atom was additionally removed from the unit cell to ensure the
electroneutrality of the crystal. In the lowest-energy structure, the Nb atom
occupies position Ti(1) (the energy of the 64-atom unit cell with Nb atom at
this position is 115~meV lower than that of the structure with Nb atom at
position Ti(2)). The conclusion about the preferential location of Nb at
position Ti(1) in KTP is consistent with the experimental
data.~\cite{SolidStateCommun.73.97}

The calculations of the second-order nonlinear susceptibility tensor show
that upon replacing titanium atoms in both positions with zirconium, the
three largest components of the tensor are reduced by $\sim$25\%
(Table~\ref{table3}) while the spontaneous polarization decreases by only 3--7\%
(to 0.240~C/m$^2$ for Zr at position Ti(1) and 0.248~C/m$^2$ for Zr at position
Ti(2)). The changes in $d_{i \nu}$ thus far exceed the changes in the spontaneous
polarization. The appearance of small components of the $d_{i \nu}$ tensor that
are missing in the $Pna2_1$ phase is due to the low symmetry of the unit cells
containing the impurity (space group $P1$).

Upon doping with niobium, the components of the second-order nonlinear
susceptibility tensor decrease more considerably, by as much as 35--40\%
(Table~\ref{table3}). Unfortunately, we are unable to correlate this effect
with the change of polarization in niobium-doped samples, since a crystal with
polarized defects such as Nb$_{\rm Ti}$--$V_{\rm K}$ has no paraelectric phase.

\begin{table}
\caption{\label{table3}Components of the second-order nonlinear susceptibility
tensor $d_{i \nu}$ in KTP crystals doped with Zr and Nb for two possible dopant
sites (in pm/V).}
\begin{ruledtabular}
\begin{tabular}{ccccccc}
Index   & $\nu = 1$ & 2 & 3 & 4 & 5 & 6 \\
\hline
\multicolumn{7}{c}{KTP : Zr(1)} \\
$i = 1$ & $-$0.691 & $-$0.548 & $-$0.805 &    0.246 &    0.693 &    0.215 \\
2       &    0.215 &    0.937 &    1.652 &    3.076 &    0.246 & $-$0.548 \\
3       &    0.693 &    3.076 &   11.099 &    1.652 & $-$0.805 &    0.246 \\
\multicolumn{7}{c}{KTP : Zr(2)} \\
1       &    0.015 &    0.243 &    0.200 & $-$0.316 &    0.658 &    0.404 \\
2       &    0.404 &    1.143 &    1.754 &    3.424 & $-$0.316 &    0.243 \\
3       &    0.658 &    3.424 &   11.712 &    1.754 &    0.200 & $-$0.316 \\
\multicolumn{7}{c}{KTP : Nb(1)} \\
1       &    1.273 &    0.824 &    1.904 & $-$0.912 &    0.741 &    0.436 \\
2       &    0.436 &    1.123 &    2.325 &    2.702 & $-$0.912 &    0.824 \\
3       &    0.741 &    2.702 &    9.767 &    2.325 &    1.904 & $-$0.912 \\
\multicolumn{7}{c}{KTP : Nb(2)} \\
1       &    0.475 &    0.104 &    0.117 &    0.064 &    1.158 &    0.237 \\
2       &    0.237 &    0.237 &    0.362 &    4.205 &    0.064 &    0.104 \\
3       &    1.158 &    4.205 &   13.936 &    0.362 &    0.117 &    0.064 \\
\end{tabular}
\end{ruledtabular}
\end{table}

In several earlier works discussing the nature of the strong optical nonlinearity
of KTP, the high values of the nonlinear optical susceptibility were associated
with strong distortion of the oxygen octahedra surrounding titanium
atoms.~\cite{ChemMater.1.492}  Our calculations of the structure of the
ferroelectric phase in doped crystals show that the distortions, which are usually
characterized by a difference between the maximum and minimum Ti--O distances
in the oxygen octahedra, are $\Delta R$(Ti(1)--O) = 0.337~{\AA} and
$\Delta R$(Ti(2)--O) = 0.214~{\AA} in undoped KTP. When Zr occupies position
Ti(1), the average values of the distortions of the TiO$_6$ octahedra are slightly
lower (0.332 and 0.207~{\AA}, respectively), whereas for Zr occupying position
Ti(2) they are slightly higher (0.342 and 0.216~{\AA}, respectively). The local
environment of the Zr impurity is characterized by substantially less distortion,
$\Delta R$(Zr(1)--O) = 0.221~{\AA} and $\Delta R$(Zr(2)--O) = 0.147~{\AA}.

As follows from our calculations, the substitution of titanium with zirconium
results in small changes in distortions of the TiO$_6$ octahedra. This contradicts
the results of X-ray measurements,~\cite{CrystallogrRep.52.659,CrystallogrRep.54.219}
which showed a marked decrease in the distortions. We believe that the reason for
this discrepancy is that the X-ray data were analyzed by assuming that the
coordinates of the oxygen atoms surrounding Ti and Zr were the same; i.e., the
difference in the local environment of these atoms was ignored.

When doping with niobium, the average values of the distortions of the TiO$_6$
octahedra are reduced to 0.305 and 0.197~{\AA}, respectively, when Nb occupies
position Ti(1); and to 0.312 and 0.215~{\AA} when Nb occupies position Ti(2).
The local environment of niobium atoms is characterized by distortions of
$R$(Nb(1)--O) = 0.272~{\AA} and $R$(Nb(2)--O) = 0.140~{\AA}. The significant
reduction of all distortions in niobium-doped sample correlates with the
strong decrease of the Curie temperature that is observed
experimentally.~\cite{JApplPhys.94.1954}

The above data show that for zirconium doping, the changes in the octahedra
distortions are too small to explain the observed changes in the nonlinear
susceptibility of doped crystals (tens of percent). For niobium doping, we
see a notable reduction in distortions of the oxygen octahedra, which also
does not explain the experimentally observed increase in the nonlinear
response of KTP at small niobium concentrations. Our results indicating
a decrease in the second-order nonlinear susceptibility when KTP is doped with
zirconium and niobium thus provide no explanation for the enhancement of the
optical nonlinearity of KTP-based solid solutions at low impurity
concentrations (at high impurity concentration, the optical nonlinearity of
crystals decreases in agreement with the results of calculations). This means
that the improvement in the nonlinear optical properties of KTP upon its
doping with Zr and Nb is not a consequence of a systematic change in the
composition of corresponding solid solutions and is likely associated with
the presence of some defects in doped samples. Such defects could be impurity
atoms substituting for other atoms besides Ti.

In niobium-doped samples, possible positions for the impurity atoms are the sites
of phosphorus atoms. Calculations of the defect formation energy according the
reaction
$${\rm KTiOPO_4} + x{\rm Nb_2O_5} \to {\rm KTiO(P_{1-2{\it x}}Nb_{2{\it x}})O_4} + x{\rm P_2O_5}$$
yield a value of 2.440~eV per impurity atom for substitution of sites P(1) and
2.368~eV for substitution of sites P(2). Interestingly, these energies are even
lower than the energies of defect formation when Nb occupies sites Ti(1) and Ti(2)
(2.490 and 2.605~eV, respectively). This means that upon doping with niobium, a
substantial part of Nb atoms can migrate to phosphorus sites, mainly sites P(2).

In zirconium-doped samples, such defects can be Zr atoms at the positions of
potassium atoms, whose coordination by oxygen atoms is 8 and 9. The existence
of such defects could explain a decrease in ionic conductivity associated with
the migration of potassium ions, which was experimentally observed in
zirconium-doped KTP. Estimates of the formation energy of such defects according
the reaction
$${\rm KTiOPO_4} + x{\rm ZrO_2} \to {\rm (Zr_{\it x}K_{1-4{\it x}})TiOPO_4} + 2x{\rm K_2O}$$
for the impurity at site K(1) yield a value of $\sim$13~eV per impurity atom.
This energy far exceeds that of defect formation when Zr occupies sites Ti(1)
and Ti(2) (0.639 and 0.484~eV, respectively) and is so high that the thermodynamic
formation of such defects is hardly possible even near the melting temperature.
However, as was established in Ref.~\onlinecite{CrystallogrRep.54.219},
the maps of the differential electron density distribution in zirconium-doped
KTP show an additional peak of the electron density near the potassium sites.
In Ref.~\onlinecite{CrystallogrRep.54.219}, this peak was explained by these
sites being populated by K atoms, but it could result from the existence of the
above defects. If this is true, the obtained data could indicate that these
defects are formed at the stage of the sample growth. At high growth rate
from a viscous solution, microinclusions of the liquid phase could be trapped
within a crystal, resulting in the formation of such nonequilibrium defects.

\begin{acknowledgments}
The calculations presented in this work were performed on a laboratory computer
cluster (16~cores) and the Chebyshev supercomputer. The author expresses his deep
gratitude to V.~I. Voronkova for her helpful discussion of the results.
\end{acknowledgments}


\providecommand{\BIBYu}{Yu}

\end{document}